\documentclass[aps,prb,twocolumn,amsmath,amssymb,superscriptaddress]{revtex4}
\usepackage{graphicx}
\usepackage{amssymb}
\usepackage{natbib}
\usepackage{color}

\newcommand{\beq}{\begin{eqnarray}}
\newcommand{\eeq}{\end{eqnarray}}

\begin{document}

\title{Flat band magnetism and helical magnetic order in Ni-doped SrCo$_2$As$_2$}
\author{Yu Li}
\email{yuli1@lsu.edu}
\affiliation{Department of Physics and Astronomy, Rice University, Houston, TX 77005, USA}
\affiliation{Department of Physics $\&$ Astronomy, Louisiana State University, Baton Rouge, LA 70803, USA}

\author{Zhonghao Liu}
\email{lzh17@mail.sim.ac.cn}
\affiliation{State Key Laboratory of Functional Materials for Informatics and Center for Excellence in Superconducting Electronics, SIMIT, Chinese Academy of Sciences, Shanghai 200050, China}

\author{Zhuang Xu}
\affiliation{Center for Advanced Quantum Studies and Department of Physics, Beijing Normal University, Beijing 100875, China}

\author{Yu Song}
\affiliation{Department of Physics and Astronomy,
Rice University, Houston, TX 77005, USA}

\author{Yaobo Huang}
\affiliation{Shanghai Synchrotron Radiation Facility, Shanghai Institute of Applied Physics, Chinese Academy of Sciences, Shanghai 201204, China}

\author{Dawei Shen}
\affiliation{State Key Laboratory of Functional Materials for Informatics and Center for Excellence in Superconducting Electronics, SIMIT, Chinese Academy of Sciences, Shanghai 200050, China}

\author{Ni Ma}
\affiliation{State Key Laboratory of Functional Materials for Informatics and Center for Excellence in Superconducting Electronics, SIMIT, Chinese Academy of Sciences, Shanghai 200050, China}

\author{Ang Li}
\affiliation{State Key Laboratory of Functional Materials for Informatics and Center for Excellence in Superconducting Electronics, SIMIT, Chinese Academy of Sciences, Shanghai 200050, China}

\author{Songxue Chi}
\affiliation{Quantum Condensed Matter Division, Oak Ridge National Laboratory, Oak Ridge, TN 37831, USA}

\author{Matthias Frontzek}
\affiliation{Quantum Condensed Matter Division, Oak Ridge National Laboratory, Oak Ridge, TN 37831, USA}

\author{Huibo Cao}
\affiliation{Quantum Condensed Matter Division, Oak Ridge National Laboratory, Oak Ridge, TN 37831, USA}

\author{Qingzhen Huang}
\affiliation{NIST Center for Neutron Research, National Institute of Standards and Technology, Gaithersburg, MD 20899, USA}

\author{Weiyi Wang}
\affiliation{Department of Physics and Astronomy,
Rice University, Houston, TX 77005, USA}

\author{Yaofeng Xie}
\affiliation{Department of Physics and Astronomy,
Rice University, Houston, TX 77005, USA}

\author{Rui Zhang}
\affiliation{Department of Physics and Astronomy,
Rice University, Houston, TX 77005, USA}

\author{Yan Rong}
\affiliation{Center for Advanced Quantum Studies and Department of Physics, Beijing Normal University, Beijing 100875, China}

\author{William A. Shelton}
\affiliation{Department of Chemical Engineering, Louisiana State University, Baton Rouge, LA 70803, USA}

\author{David P. Young}
\affiliation{Department of Physics $\&$ Astronomy, Louisiana State University, Baton Rouge, LA 70803, USA}

\author{J. F. DiTusa}
\email{ditusa@phys.lsu.edu}
\affiliation{Department of Physics $\&$ Astronomy, Louisiana State University, Baton Rouge, LA 70803, USA}

\author{Pengcheng Dai}
\email{pdai@rice.edu}
\affiliation{Department of Physics and Astronomy,
Rice University, Houston, TX 77005, USA}
\affiliation{Center for Advanced Quantum Studies and Department of Physics, Beijing Normal University, Beijing 100875, China}

\date{\today}

\begin{abstract}
A series of Sr(Co$_{1-x}$Ni$_x$)$_2$As$_2$ single crystals was synthesized allowing a comprehensive phase diagram with respect to field, temperature, and chemical substitution to be established. Our neutron diffraction experiments revealed a
helimagnetic order with magnetic moments ferromagnetically (FM) aligned in the $ab$ plane and a helimagnetic wavevector of $q=(0,0,0.56)$ for $x$ = 0.1. The combination of neutron diffraction and angle-resolved photoemission spectroscopy (ARPES) measurements show that the tuning of a flat band with $d_{x^2-y^2}$ orbital character drives the helimagnetism and indicates the possibility of a quantum order-by-disorder mechanism.
\end{abstract}

\maketitle

\section{INTRODUCTION}
The exploration of materials in proximity to quantum phase transitions is a fruitful
area for discovering exotic states of matter due to the strong influence of quantum fluctuations. This idea can be traced to Villain who coined the name \textit{order-by-disorder} when investigating frustrated insulating magnets \cite{Villain1980}.
The central idea refers to the fact that quantum fluctuations, akin to classical entropic effects, can lift the large degeneracy of ground states and favor a particular order, hence the term {\it order-by-disorder} \cite{Green}. This is considered to be the quantum equivalent to the theme used to explain a variety of phenomena including surface tension and DNA folding, and has been employed to explain unconventional superconductivity \cite{Fay1980} and even emergent gravity \cite{Verlinde}. There is large recent theoretical interest in itinerant ferromagnets in which quantum fluctuations dominate and drive the system into a variety of novel magnetic states in the vicinity of a quantum critical point \cite{Conduit,Uhlarz2004,Thomson}, such as a helical state driven by ferromagnetic (FM) spin fluctuations. Meanwhile, as an origin of ferromagnetism complementary to that described by Nagaoka \cite{Nagaoka}, flat-band physics \cite{Tasaki} has garnered attention since it provides fertile ground for diverse interaction-driven quantum phases including FM order, a Mott insulating phase \cite{Cao1}, and superconductivity \cite{Cao2}. Therefore, it is interesting to search for helimagnetism associated with a flat-band lying close to the chemical potential in a material close to a quantum critical point.

In the $A$Fe$_2$As$_2$ ($A=$Ca,Sr,Ba) family of iron pnictides, it is well known that the superconducting electron pairing is attributed to antiferromagnetic (AF) spin fluctuations \cite{scalapinormp,Dai_RMP}. However, FM spin fluctuations may also be important to electron pairing \cite{Singh2008,Mazin2008} and were recently observed in Co-substituted compounds by nuclear magnetic resonance (NMR) \cite{Wiecki} and neutron scattering experiments \cite{Yu_SrCo2As2,Sapkota2017}. Furthermore, electronic structure calculations \cite{Pandey2013,Mao2018} and angle resolved photoemission spectroscopy (ARPES) experiments indicate that $A$Co$_2$As$_2$ \cite{Yu_SrCo2As2,Xu2013,Sefat2009} is in proximity to a FM instability due to the existence of a flat band nearby the Fermi level, although these materials remain paramagnetic down to 2 K with AF low energy spin fluctuations \cite{Yu_SrCo2As2}. Since chemical substitution can shift the Fermi level relative to the band structure \cite{Liu2018}, substitution of Co with Ni will drive the system toward a Van Hove singularity associated with the flat band and efficiently promote the FM instability. Therefore, it is interesting to investigate how AF and FM spin fluctuations evolve in $A$(Co$_{1-x}$Ni$_x$)$_2$As$_2$ and to explore the relevant emergent phenomena.

We successfully synthesized single crystals [Fig.1(a)] of a series of Sr(Co$_{1-x}$Ni$_x$)$_2$As$_2$ with a range of Ni concentrations between 0 and 0.6, and carried out systematic magnetic susceptibility, $\chi$, and magnetization, $M$, measurements. Sr(Co$_{1-x}$Ni$_x$)$_2$As$_2$ has a ThCr$_2$Si$_2$-type body-centered tetragonal crystal structure with no indication of a structural transition [Fig. 1(b)] \cite{SI}. Considering both SrCo$_2$As$_2$ \cite{Pandey2013,Yu_SrCo2As2} and SrNi$_2$As$_2$\cite{SrNi2As2} are paramangetic without magnetic order at low temperature, it is surprising that we discovered a helical magnetic order in Sr(Co$_{0.9}$Ni$_{0.1}$)$_2$As$_2$ with a magnetic wave vector $q$ = (0,0,0.56) and transition temperature $T_c = 28$ K [Fig. 1(c) and 2(c)]. We find that the transition temperature varies systematically with substitution and has maximum value at $x = 0.1$ from susceptibility measurements [Fig. 2(c)]. Our results suggests that the ground state is helimagnetic (HM) with moments lying parallel to the $ab$ easy plane while rotating with respect to the $c$-axis in adjacent layers [Fig. 1(c)]. Measurements of $M(H)$ with an in-plane magnetic field at low temperatures (2 K $\leq T \leq$ 4 K) demonstrate a two-step transition from HM order first into a partially polarized magnetic (PPM) phase and ultimately into a fully polarized magnetic (FPM) state. Under a field parallel with the c-axis, the magnetization is linearly dependent on $H$ prior to saturation at the higher field. Combining neutron diffraction and ARPES experiments, we established a close connection between the magnetic moments and a flat band of the $d_{x^2-y^2}$ orbital character. Since it was suggested that BaCo$_2$As$_2$ is a quantum paramagnet near a FM critical point \cite{Sefat2009,Nakajima2019}, the helimagnetic order observed in Ni-substituted SrCo$_2$As$_2$ may be induced by quantum fluctuations in the vicinity of the putative quantum critical point \cite{Green}.

\begin{figure}[htb]
\includegraphics[scale=.45]{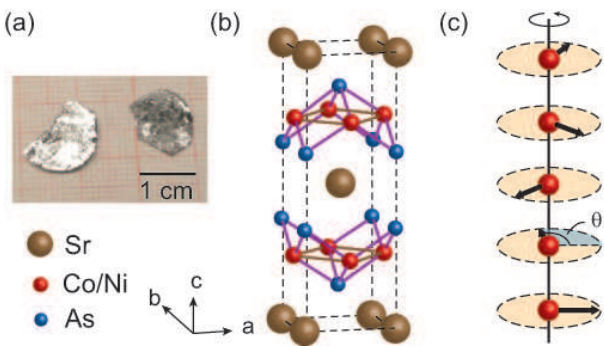}
\caption[Fig1]{(a) Single crystals of Sr(Co$_{1-x}$Ni$_x$)$_2$As$_2$. (b) Tetragonal structure of SrCo$_2$As$_2$. (b) Helical magnetic structure along the $c$-axis with a rotation angle $\theta$ between neighboring layers. }
\end{figure}

\section{EXPERIMENTAL DETAILS}
\subsection{ Sample Growth and Magnetization measurements}

Single crystals of Sr(Co$_{1-x}$Ni$_x$)$_2$As$_2$ were grown from solution using a self-flux method with the ratio Sr:(Co$_{1-x}$Ni$_x$):As = 1:5:5. The elements were placed in an aluminum oxide crucible and sealed in an evacuated quartz tube. After heating slowly below 830$^\circ$C, the mixture was cooked at 1200$^\circ$C for 20 hours and then slowly cooled down to 1050$^\circ$C at a rate of 3$^\circ$C/h and then down to 800$^\circ$C at 10$^\circ$C/h. Single crystals were obtained by cleaning off the flux. The typical crystals were about 1 cm$^2$ in size and a few mm in thickness [Fig. 1(a)].

Temperature- and field-dependent dc magnetization measurements were performed on a Quantum Design (QD) Physical Property measurement system (PPMS) in Beijing Normal University and a Magnetic Property Measurement System (MPMS) in Louisiana State University.

\begin{figure}[thb!]
\includegraphics[scale=.45]{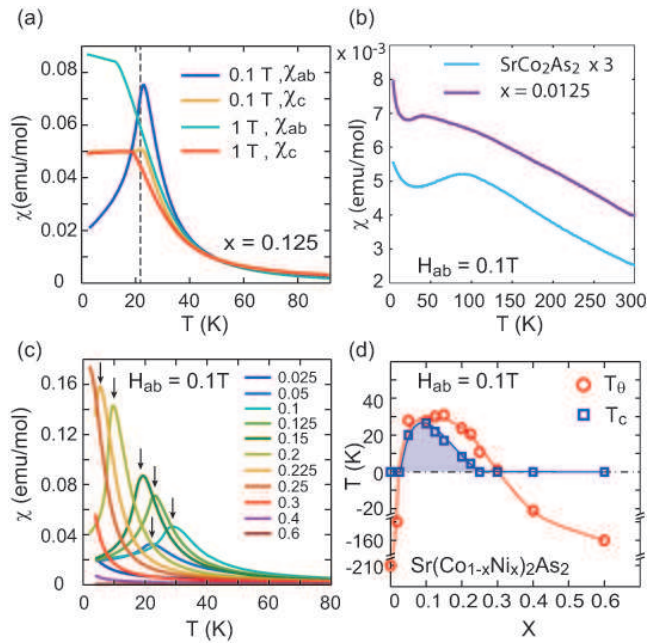}
\caption{(a) Magnetic susceptibility, $\chi$, in Sr(Co$_{1-x}$Ni$_x$)$_2$As$_2$ with field $H = 0.1$ and 1-T parallel and perpendicular to $c$-axis for $x$ = 0.125. (b) and (c) $\chi(T)$ for crystals with $0 \le x \le 0.6$. (c) Critical temperature, $T_c$, and Weiss temperature $T_{\theta}$ $\textit{vs.}$ $x$ in Sr(Co$_{1-x}$Ni$_x$)$_2$As$_2$.
}
\end{figure}

\subsection{ Neutron Diffraction}

Neutron powder diffraction (NPD) data for compound Sr(Co$_{0.9}$Ni$_{0.1}$)$_2$As$_{2}$ were collected at the NIST Center for Neutron Research on the high resolution powder neutron diffractometer (BT-1) with neutrons of wavelength 1.5403 {\AA} (at 295 K) and 2.0775 {\AA} (at 30 and 5 K) produced by using Cu(311) and Ge(311) monochromators, respectively.  Collimators with horizontal divergences of 15$'$, 20$'$ and 7$'$ of arc were used before and after the monochromator and after the sample, respectively.  Data were collected in the 2$\theta$ range of 3-168$^\circ$ with a step size of 0.05$^\circ$.  The structural parameters were refined by Rietveld refinement using the GSAS and EXPGUI programs \cite{Toby,Larson}. The atomic neutron scattering factors used in the refinements for Sr, Co, Ni, and As were 0.702, 0.253, 1.03, and $0.658 \times10^{-12}$ cm, respectively. About 5 grams of Sr(Co$_{0.9}$Ni$_{0.1}$)$_2$As$_2$ single crystals were grinded into powder for the neutron powder diffraction measurements. Single crystal neutron diffraction measurements were carried out on two instruments at the High Flux Isotope Reactor (HFIR) at Oak Ridge National Laboratory (ORNL), the single crystal diffractometer HB3A and the recently upgraded wide angle neutron diffractometer (WAND2) at the HB-2C beam port. The monochromatic incident beam of wavelength 1.546 \AA\ was used on HB3A diffractometer which is provided by a Si(220) monochromator\cite{Chakoumakos}. The detector on HB3A is a two-dimensional (2D) scintillation Anger camera with 1 mm spatial resolution. WAND2 uses a monochromatic beam with a wavelength of 1.48 \AA\ provided by the Ge(311) monochromator. It is equipped with a 2D position sensitive detector (PSD), which covers the scattering angles of 120$^\circ$ and $\pm$7.5$^\circ$ in the horizontal and vertical directions, respectively. An oscillating radial collimator is used before the PSD \cite{Frontzek} On both diffractometers the sample was inserted into a Closed Cycle Refrigerator that provided a temperature range from 4 K to 300 K.

\begin{figure}[thb!]
\includegraphics[scale=.45]{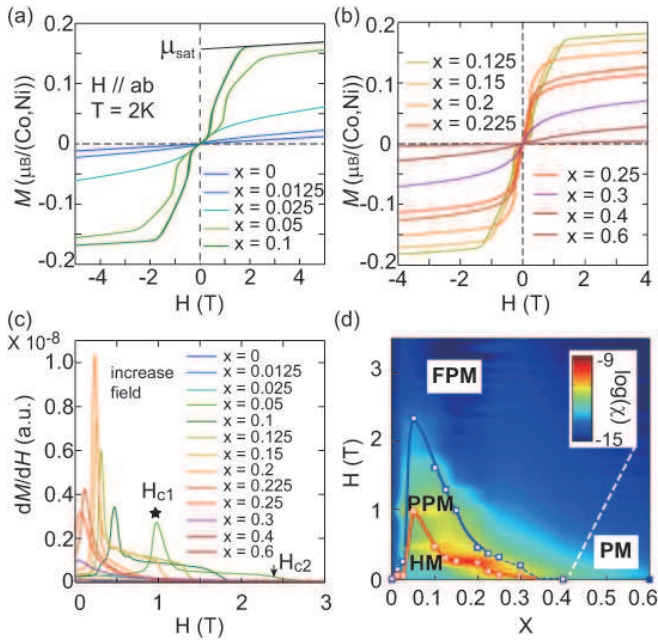}
\caption{(a),(b) M(H) for crystals with $0 \le x \le 0.6$. (c) $dM/dH$ $\textit{vs.}$ $H$ for $0 \le x \le 0.6$. (d) Magnetic phase diagram with $H$ and $x$. $H_{c1}$ and $H_{c2}$ are defined in frame (c). $H_{c2}$ is  represented as blue dashed curve where it is poorly determined by the data. The white dashed line is a speculative separation (or crossover) between paramagnetic (PM) and field- induced FM states. PPM refers to the partially polarized magnetic state, and FPM means fully polarized magnetic state.
}
\end{figure}

For single crystal diffraction measurements on Sr(Co$_{0.9}$Ni$_{0.1}$)$_2$As$_2$, we define the momentum transfer ${\bf Q}$ in three-dimensional reciprocal space in {\AA}$^{-1}$ as $\textbf{Q}=H\textbf{a}^\ast+K\textbf{b}^\ast+L\textbf{c}^\ast$, where $H$, $K$, and $L$ are Miller indices and
${\bf a}^\ast=\hat{{\bf a}}2\pi/a$, ${\bf b}^\ast=\hat{{\bf b}}2\pi/b$, ${\bf c}^\ast=\hat{{\bf c}}2\pi/c$ with  $a= b\approx 3.95$ {\AA} , and $c=11.62$ {\AA}.

\subsection{ Angle-resolved Photoemission Spectroscopy (ARPES)}

\begin{figure}[thb!]
\includegraphics[scale=.6]{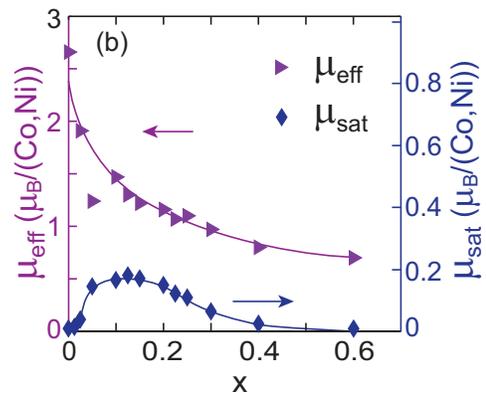}
\caption{Effective magnetic moment, $\mu_{eff}$, by fits of the C-W law to the magnetic susceptibility above $T_c$, and saturated magnetic moment, $\mu_{sat}$, $\textit{vs.}$ $x$. The large ratio of $\mu_{eff}/\mu_{sat}$ reflects the itinerant nature of magnetism.
}
\end{figure}

ARPES measurements were performed at the Dreamline beam line of the Shanghai Synchrotron Radiation Facility using a Scienta DA80 analyzer and at the beam line 13U of the National Synchrotron Radiation Laboratory (Hefei) equipped with a Scienta R4000 analyzer. The energy and angular resolutions were set to 15 meV and 0.2$^\circ$, respectively. Samples were cleaved $\textit{in situ}$, yielding flat mirror-like (001) surfaces. During the measurements, the temperature was kept at 20 K and the pressure was maintained better than $5\times 10^{-11}$ Torr.

\section{RESULTS}

\subsection{Magnetization and magnetic susceptibility data}

The magnetic susceptibility $\chi(T)$ measured for Sr(Co$_{1-x}$Ni$_x$)$_2$As$_2$ with $x = 0.125$ is shown in Fig. 2(a). We find a clear magnetic phase transition at $T_c = 22$ K, characterized by a maximum in $d(\chi \cdot T)/dT$ \cite{fisher1962} for $H=1$ kOe and $H$ both parallel and perpendicular to the $c$-axis. For $H = 0.1$ T, $\chi(T < T_c)$ displays a strong dependence on the direction of $H$ relative to the crystalline lattice, exhibiting a behavior typical of  AF order. In contrast to the $A$-type AF order in CaCo$_2$As$_2$ where spins are ordered along the $c$-axis, the anisotropy of $\chi (T)$ for Sr(Co$_{1-x}$Ni$_x$)$_2$As$_2$ clearly indicates that while the system is isotropic at high temperature, the spins are aligned in $ab$ plane below $T_c$\cite{SI}, exhibiting the typical behavior of AF magnet with an easy plane. By fitting the high temperature susceptibility (80$-$200 K) with the Curie-Weiss law [$\chi(T) \propto (T-T_{\theta})^{-1}$], we find a positive $T_{\theta}$ ($\approx 28$ K for $H_{ab}$ and 25K for $H_c$) \cite{SI}, suggesting that the system is dominated by FM interactions.

\begin{figure}[thb!]
\includegraphics[scale=.4]{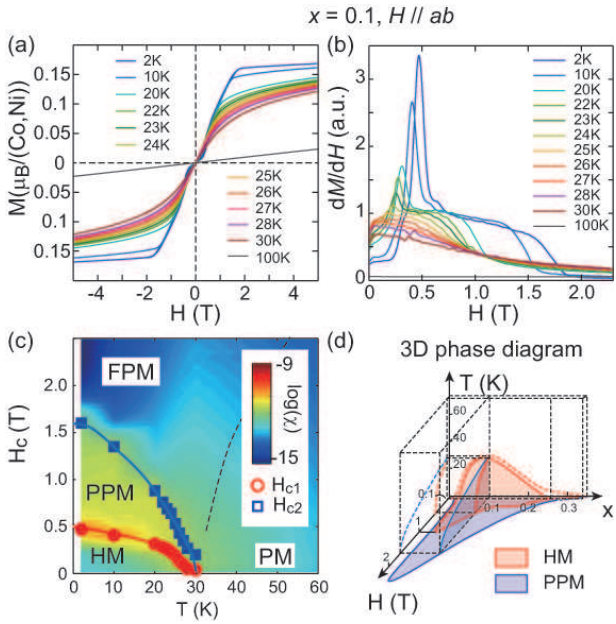}
\caption{(a),(b) $M_{ab}$ and $dM_{ab}/dH$ $\textit{vs.}$ $H$ at temperatures identified in the figures. (c) Phase diagram determined from data in frames (a) and (b). Dahsed line is a speculative crossover separating the fully polarized magnetic, FPM, and paramagnetic, PM, phases. PPM denotes the partially polarized magnetic phase. (d) Three dimensional phase diagram combining Fig.2(d), 3(d) and 7(c).
}
\end{figure}

In classical molecular-field theory \cite{Johnston2016}, the parameter $f \equiv T_{\theta}/{T_c}$ is usually found to be smaller than 1 with $-\infty < f < -1$ for antiferromagnets and 1 for ferromagnets. This is because the real transition (either FM or AF) has to arise before the divergence of $\chi(q)$ which describes the instability of the system towards a particular magnetic state. Therefore, the divergence of Curie-Weiss susceptibility at $T_{\theta} > T_c$ ($f > 1$) reflects that the system is near a FM instability. However, a suppression of the uniform susceptibility [$\chi(q=0)$] above $T_c$ indicates the influence of low dimensionality since a 2D isotropic Heisenberg system
cannot exhibit magnetic order above $T = 0$. This strongly suggests quantum fluctuations may play a significant role. In addition, the large enhancement of $\chi_{ab}(T < T_c)$ with in-plane field at $H = 1$ T suggests the existence of a field-induced meta-magnetic transition.

In order to understand the evolution of this magnetic state with $x$, we systematically explore $\chi(T)$ for Sr(Co$_{1-x}$Ni$_x$)$_2$As$_2$ single crystals. In Fig. 2(b), we show magnetic susceptibility of stoichiometric SrCo$_2$As$_2$ and slightly Ni-doped Sr(Co$_{1-x}$Ni$_x$)$_2$As$_2$ with $x = 0.0125$. There is a broad maximum arising  around 100 K for $x = 0$ and 50 K for $x = 0.0125$, similar to previous report on SrCo$_2$As$_2$\cite{Pandey2013} in which the maximum in $\chi(T)$ was attributed to a cross-over from a coherent to an incoherent Fermi liquid state with increasing $T$. However, we note that this broad maximum also exists in other itinerant magnetic materials in the proximity to ferromagnetism, such as CrGe \cite{Klotz}. The gradual suppression with Ni substitution might suggest its close relationship with the novel magnetic phase observed in compounds with larger Ni substitution. With larger $x$ as shown in Fig.~2(c), we see that the magnetic transition temperature depends on $x$ in a systematic way creating a dome-like feature in the $x$-$T$ phase diagram with a maximum $T_c$ at $x \sim 0.1$ [Fig.~2(d)].  In Fig.~2(d), we summarize $T_{\theta}$ by fitting the Curie-Weiss form to $\chi(T)$ for $0\le x \le 0.6$\cite{SI}. Curiously, for Ni concentrations within the dome of magnetic ordering, the dominant interaction, as assessed by $T_{\theta}$, is ferromagnetic ($T_{\theta}>0$) with $f > 1$, while an AF interaction ($T_{\theta} <0)$ dominates outside this dome. The concurrence of the HM phase transition and the positive $T_{\theta}$ strongly indicates that the magnetic order is driven by FM interactions.

\begin{figure}[thb!]
\includegraphics[scale=.45]{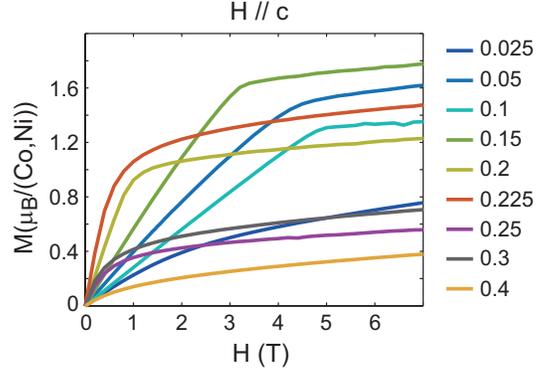}
\caption{Magnetization,$M$, $\textit{vs.}$ magnetic field, $H$ with $H$ parallel to the c-axis at a variety of Ni concentrations.
}
\end{figure}

In Figs.~3(a) and (b), we applied a magnetic field along the $ab$-plane and measured $M(H)$ for a series of Ni concentrations. Field-induced meta-magnetic transitions associated with the reorientation of magnetic moments are clearly observed in the magnetically ordered compounds. In Fig.~3(a), for example, at $x=0.1$, the system is in a helical magnetic ground state at low temperatures. The magnetization displays a linear dependence on the in-plane magnetic field at low field [Fig. 3(c)]. As the field increases, a partially polarized magnetic (PPM) intermediate state arises at $H_{c1} \sim 0.5$ T. This transition is clearly observed in $\chi(H) \equiv dM/dH$ as a sharp peak which defines the critical field as $H_{c1}$ [Fig. 3(c)]. The magnetization around $H_{c1}$ exhibits hysteresis behavior\cite{SI} with magnetic field, implying a first-order transition. In the intermediate state, the susceptibility displays a plateau between $H_{c1}$ and $H_{c2}$ beyond which the system evolves into a fully polarized magnetic (FPM) state. Furthermore, as the Ni concentration changes, the plateau in $\chi(H)$ shrinks and disappears for $x < $ 0.05 or $x \geq$ 0.2, such that the determination of $H_{c2}$ in these regions is somewhat ambiguous. Therefore, we employed a blue dashed line in the $x$-$H$ phase diagram to represent the poorly defined $H_{c2}$ [Fig. 3(d)]. We have included the magnetic susceptibility ($\chi \equiv dM/dH$) as a color plot (log scale) in Fig. 3(d) to emphasize the clear delineation of magnetic phases.

\begin{figure}[thb!]
\includegraphics[scale=.45]{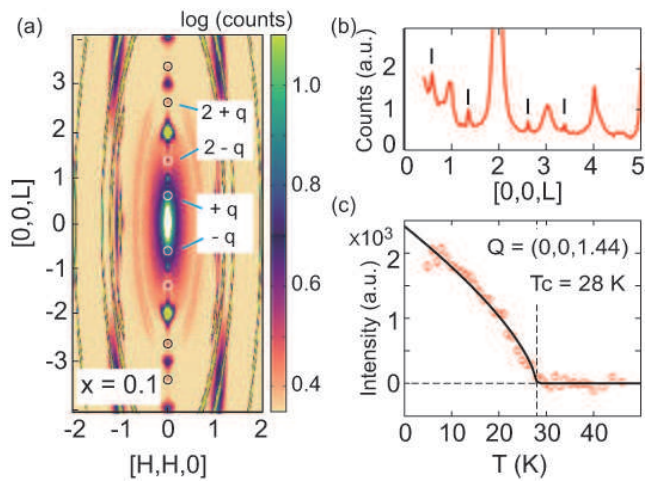}
\caption{(a) Neutron diffraction measurements on Sr(Co$_{0.9}$Ni$_{0.1}$)$_2$As$_2$ in the $[H,H,L]$ plane \cite{SI}. (b) One-dimensional cut along the $[0,0,L]$ direction in (a). The intensities at (0,0,L) with L = odd number are due to double scattering\cite{doube_scattering}. (c) Temperature dependence of the magnetic peak at $Q=(0,0,1.44)$.
}
\end{figure}

In Fig. 4, the effective magnetic moment, $\mu_{eff}$, estimated from the Curie constant, and the saturated magnetic moment, $\mu_{sat}$, according to the high field magnetization, are presented. The magnitude of the effective moment is similar to that for an S=1/2 system while $\mu_{sat}$ is about one order of magnitude smaller for $x > 0.05$. This renders a large Rhodes-Wholfarth ratio, $\mu_{eff}$ / $\mu_{sat}$, suggesting that the magnetism in this system is itinerant in character.

\begin{figure}[thb!]
\includegraphics[scale=.6]{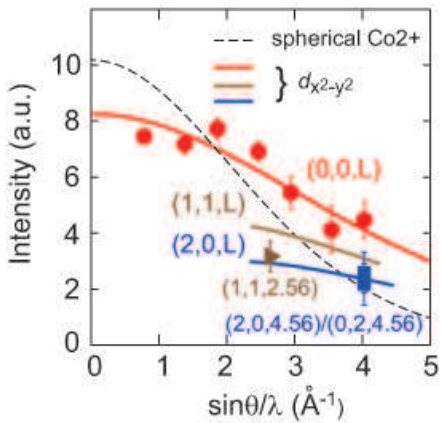}
\caption{Integrated intensity of magnetic Bragg peaks. The dashed curve is the magnetic form factor of Co$^{2+}$ using a spherical approximation. The solid curves are magnetic form factors for the $d_{x^2-y^2}$ orbital multiplied with a polarization factor along different measured directions for a helical order. Red, brown and blue are along the $[0,0,L]$, $[1,1,L]$, and $[2,0,L]/[0,2,L]$ directions, respectively.
}
\end{figure}

In Figs. 5(a) and (b), we investigated the temperature dependence of $M(H)$ and $\chi(H)$ for Sr(Co$_{1-x}$Ni$_x$)$_2$As$_2$ with $x =$ 0.1. The plateau in $\chi(H)$ was observed at all temperatures below $T_c$ [Fig.5(b)]. We note that even in the paramagnetic state, the magnitude at $T$ = 30 K is dramatically enhanced as the system approaches the magnetic transition temperature, in comparison with that at a higher temperature of $T = 100$ K. Based on these measurements, a $H$-$T$ phase diagram for $x=0.1$ is presented in Fig.~5(c) and a schematic three-dimensional ($H$-$T$-$x$) phase diagram is shown in Fig.~5(d). In Fig. 6, we plot the magnetization of Sr(Co$_{1-x}$Ni$_x$)$_2$As$_2$ over a range of $x$ with field parallel to the $c$-axis. The magnetization at low fields has a linear dependence on field, before saturating at higher fields. In summary, our magnetization and susceptibility measurements on Sr(Co$_{1-x}$Ni$_x$)$_2$As$_2$ are consistent with the magnetic behavior of a system with helical spin order. However, further neutron diffraction measurements are required to determine the magnetic structure of the ground state.

\begin{figure}[htb]
\includegraphics[scale=.62]{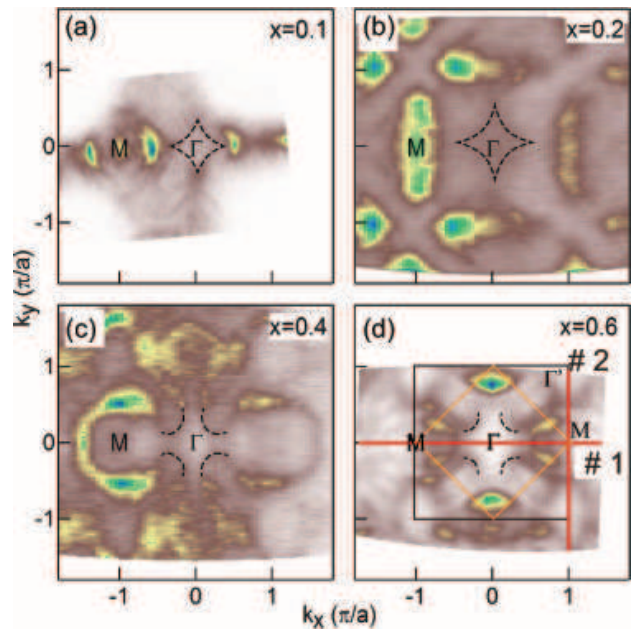}
\caption{ Plot of the ARPES intensity at $E_F$ of Sr(Co$_{1-x}$Ni$_{x}$)$_2$As$_2$ with x = 0.1, 0.2, 0.4, and 0.6. A complex structure of Fermi surfaces is observed and the topology changes dramatically from x=0.1 to x=0.6 due to the existence of flat band dispersion. The dashed lines curves are guide to the eye.
}
\end{figure}

\subsection{Neutron diffraction}

\begin{figure*}[htb]
\includegraphics[scale=.7]{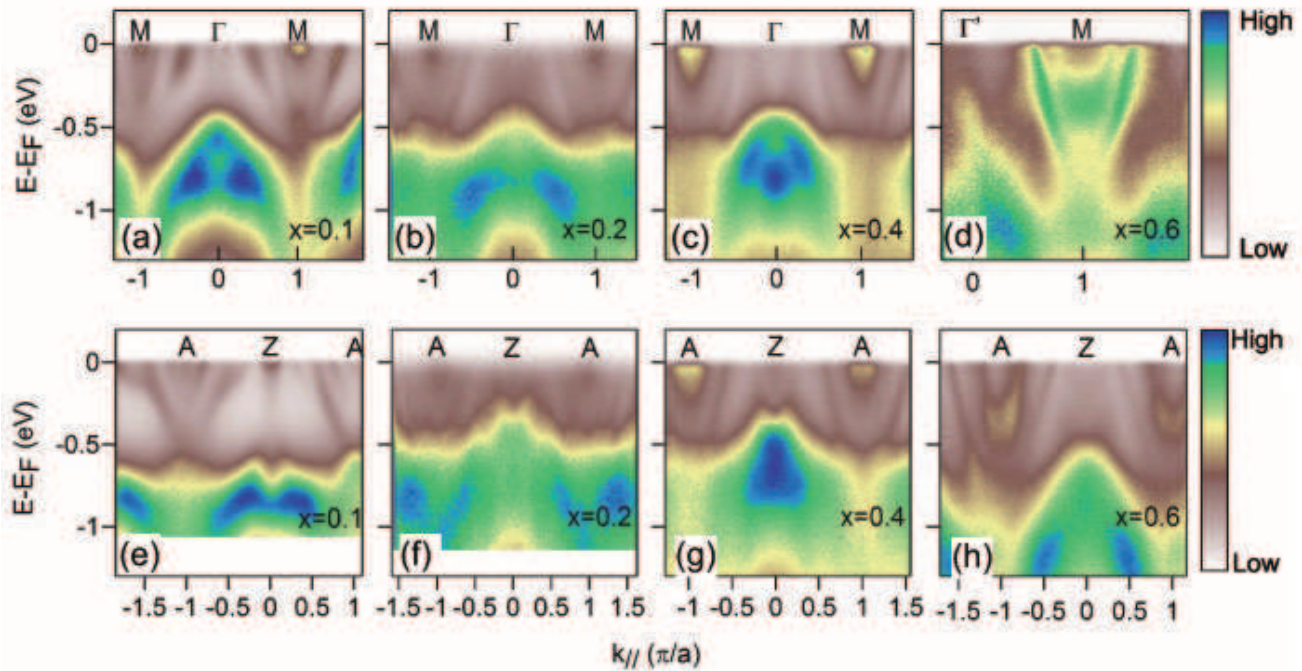}
\caption{Intensity plots of the band dispersion along the$\Gamma$-M (a-d) and Z-A (e-h) directions from ARPES measurements for $x$ = 0.1, 0.2, 0.4, and 0.6.
}
\end{figure*}

In order to understand the changes that occur with $x$ in Sr(Co$_{1-x}$Ni$_x$)$_2$As$_2$, we probed the magnetic structure at zero field and $T=4$ K for $x = 0.1$ via neutron diffraction experiments. We chose to explore a Sr(Co$_{0.9}$Ni$_{0.1}$)$_2$As$_2$ single crystal as these samples displayed the highest $T_c$ with one of the largest $\mu_{sat}$ for this series. Figure~7(a) displays the neutron diffraction pattern in the $[H,H,L]$ scattering plane. Incommensurate peaks at $(0,0,n\pm q)$ with $n$ an even integer and $q = (0,0,0.56)$ are clearly observed in Fig~7(b), manifesting either a planar spiral or sinusoidal magnetic structure. However, a sinusoidal magnetic structure is unlikely in this material since it requires a uniaxial spin anisotropy along either the $a$ or $b$ axis and therefore breaks the tetragonal symmetry of the underlying crystalline lattice\cite{SI}. This is substantially different from the collinear AF order in iron pnictides \cite{Dai_RMP} or the commensurate $A$-type AF order in CaCo$_2$As$_2$ \cite{Quirinale2013}.
In EuCo$_2$As$_2$ \cite{EuCo2P2}, a similar helical order has been observed. However, the magnetic moments are localized on Eu sites with $S = 7/2$, in contrast to the itinerant magnetism in Sr(Co$_{1-x}$Ni$_x$)$_2$As$_2$, implying the underlying physics is significantly different. In addition, the dissimilar spin anisotropy between CaCo$_2$As$_2$ \cite{Zhang2015} and Sr(Co$_{1-x}$Ni$_x$)$_2$As$_2$ suggests a change from an easy-axis to an easy-plane spin anisotropy\cite{Bing2019}. Furthermore, the value of $q$ is close but not equal to $(0,0,0.5)$ indicating a short pitch for the helix, about 4 Co-layers or 2 unit cells, resulting in spins on adjacent Co-layers being nearly perpendicular. This severely constrains the likely exchange interaction responsible for the magnetism. Figure~7(c) demonstrates the $T$-dependence of the scattering intensity at $Q = (0,0,1.44)$. The $T_c$ of 28 K agrees well with that determined from $\chi(T)$ in Fig.~2(c). A longitudinal scan along the $[0,0,L]$ direction at 4 K (Fig. S4 in Ref. \cite{SI}) indicates that the magnetic ordering is long-range within the current instrumental resolution.

In Fig.~8, we show the magnetic intensity from neutron diffraction measurements in comparison with the calculated structure factor for helical order. We employed a $d_{x^2-y^2}$ magnetic form factor for Co$^{2+}$ ions \cite{mmf,SI} and the results (solid lines) are in good agreement with our neutron diffraction measurements. This agreement suggests that the magnetic moments in Sr(Co$_{1-x}$Ni$_x$)$_2$As$_2$ are closely associated with electrons having $d_{x^2-y^2}$ orbital character. The spherical Co$^{2+}$ magnetic form factor (dashed curve ) is plotted for comparison and is a much poorer representation of our data.

\subsection{Electronic band structure}

In nominally pure SrCo$_2$As$_2$, a flat band exists at an energy just above the Fermi level near the $M$ point of the Brillouin Zone (BZ) \cite{Mao2018,Yu_SrCo2As2}. Substitution of Ni is expected to raise $E_F$ so that this flat band is partially occupied resulting in $E_F$ being closer to a Van Hove singularity. To establish the connection between the helical magnetic phase and this flat band, we carried out ARPES experiments on a series of Sr(Co$_{1-x}$Ni$_x$)$_2$As$_2$ samples ($x = $0, 0.1, 0.2, 0.4 and 0.6) to map out their electronic structures. We show in Fig.~9 the Fermi surfaces of Sr(Co$_{1-x}$Ni$_x$)$_2$As$_2$ with $x =$ 0.1, 0.2, 0.4, 0.6 as measured by ARPES. A diamond-shaped electron-like Fermi surface (dashed curve) is observed at the $\Gamma$ point in the sample with $x = 0.1$, similar to that observed in nominally pure  Sr(Ba)Co$_2$As$_2$ \cite{Xu2013,Yu_SrCo2As2}. This band is associated with a $d_{x^2-y^2}$ orbital and has flat dispersion. It is likely responsible for in-plane ferromagnetism and the observed helimagnetic order. We note that as the Ni concentration increases, this Fermi surface opens along the $\Gamma$-$M$ direction, and is further separated at $x$ = 0.6, consistent with the fact that the flat band shifts down below the Fermi level.

\begin{figure}[htb]
\includegraphics[scale=.6]{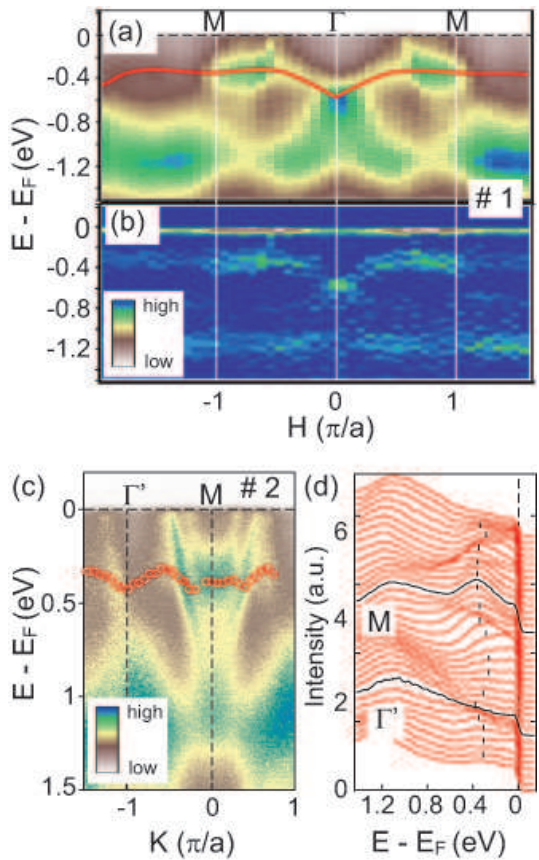}
\caption{(a) and (b) ARPES Intensity plot and its second derivative along the cut $\#$1 direction as shown in Fig. 9 (d). (c) and (d) ARPES spectra along the cut $\#$2 direction and its corresponding Energy distribution curve (EDC).
}
\end{figure}

In Fig.~10, we display the cuts of ARPES spectra along the $\Gamma$-$M$ and $Z$-$A$ directions for these samples. While there are barely changes in in $x$ = 0.1, 0.2 and 0.4 samples, the whole band structure suddenly shifts down for $x = 0.6$. This shows that the flat band along the $\Gamma$-$M$ direction sinks below the $E_F$ and thus is only observed in ARPES spectra of Fig. 10(d). In order to further prove the existence of the flat band, we made two cuts along both the horizontal and vertical directions labeled as red lines in Fig. 9(d) (for $x$ = 0.6) and display these in Fig. 11. Figures~11(a) and (c) are the intensity plots of band dispersion along the cut $1$ and cut $2$ directions. A nearly flat band dispersion around $-0.4$ eV with an electron-like band minimum at the $\Gamma$ point is realized in both cuts. These flat band spectra typically have low intensity in ARPES \cite{Fe3Sn2}, but are clearly resolved in the second-derivative plot [Fig.11(b)] and the energy distribution curves (EDCs) [Fig.11(d)].

To investigate the $x$-dependence of this flat band, we show the EDC cuts at the $M$ point for samples with $0\leq x \leq 0.6$ in Fig.~12(a). One peak close to $E_F$ associated with the flat band is identified by the arrows and is degenerate with another electronic band at the $M$ point \cite{Mao2018,Yu_SrCo2As2}. The Fermi-Dirac function cuts off ARPES spectra at $E_F$ affecting the shape of these peaks. To avoid this effect, we use the symmetrized EDCs to determine whether the band is located below or above the Fermi energy. Similar methods have been applied to determine small superconducting gaps in iron based superconductors\cite{Yu_LiFeAs} [Fig. 12(b)]. A central peak at $E_F$ manifests that the band is either at or above the Fermi energy and thus gives the lower limit of the band location. We use the value estimated from dynamic mean field theory (DMFT) calculations on pure SrCo$_2$As$_2$ in previous work \cite{Mao2018,Yu_SrCo2As2} as the upper limit and the results are summarized in Fig.~12(c). The shading area roughly establishes the evolution of the energy location of the flat band with Ni substitution. It is clear that the band shifts with $x$ at $x < 0.4$ much more slowly than at larger $x$. This is a manifestation of the large density of states (DOS) of the flat band for electrons filling as it shifts across $E_F$ ($0<x<0.4$). This flat band has $d_{x^2-y^2}$ orbital character and makes a significant contribution to DOS near the Fermi energy \cite{Yu_SrCo2As2}. Our ARPES results for $0\leq x\leq 0.6$ confirm that this flat band is partially occupied for the range of $x$ with helimagnetism, and suggest that the helical magnetic phase is closely associated with this band.

\begin{figure}[htb]
\includegraphics[scale=0.35]{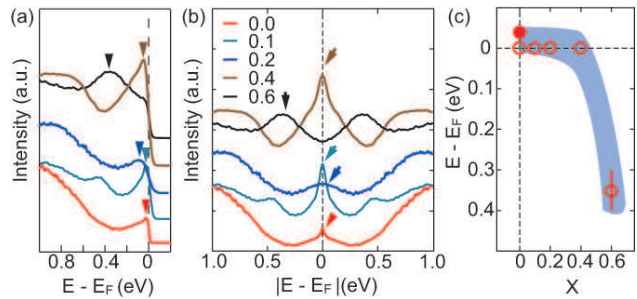}
\caption{(a) Energy distribution curves (EDCs) at the M point for Sr(Co$_{1-x}$Ni$_x$)$_2$As$_2$ with $x$ = 0, 0.1, 0.2, 0.4, 0.6. Small peaks close to $E_F$ were identified as residual spectra from the band above the Fermi level truncated by Fermi-Dirac function. (b) Symmetrized EDCs. The peaks at E$_F$ in symmetrized EDCs set the lower limit of the energy location of the flat band within the instrumental resolution. (c) Energy location of the flat band determined by ARPES and that from DMFT calculation\cite{Mao2018} for $x$ = 0 which we used as the upper limit. The shaded area is a guide to the eye. The evolution of the energy of the flat band with respect to the Fermi level is demonstrated.
}
\end{figure}

\section{DISCUSSION AND CONCLUSION}

In $A$Co$_2$As$_2$ systems, density functional theory (DFT) plus DMFT calculations suggest that FM instabilities are due to a flat band residing near the $E_F$. However, experiments show an evolution from AF order in CaCo$_2$As$_2$ \cite{Quirinale2013,Sapkota2017} to paramagnetism in BaCo$_2$As$_2$\cite{Sefat2009} (or Ba(Fe$_{1/3}$Co$_{1/3}$Ni$_{1/3}$)$_2$As$_2$ \cite{Nakajima2019}) with evidence for a nearby FM  critical point \cite{Sefat2009,Nakajima2019}. This indicates that the magnetism can be tuned by chemical substitution. In SrCo$_2$As$_2$, previous inelastic neutron scattering experiments \cite{Yu_SrCo2As2} revealed the coexistence of FM and AF spin fluctuations. It is likely that the fine tuning of these fluctuations by chemical substitution induces the observed magnetic order in Sr(Co$_{1-x}$Ni$_x$)$_2$As$_2$.

In many helical magnetic systems such as MnSi \cite{Dhital2017,Muhlbauer2009,MnSi1} or Cr$_{1/3}$NbSe$_2$ \cite{Cr2,Ghimire2013}, the helical magnetic order is induced by the Dzyaloshinskii-Moriya (DM) interaction \cite{DM} arising from the lack of crystalline inversion symmetry. Since Sr(Co$_{1-x}$Ni$_x$)$_2$As$_2$ has a centro-symmetric crystal structure, the DM interaction is absent.In isostructural EuCo$_2$As$_2$ where a large mangetic moment exists on the Eu site\cite{Johnston2016}, the mechanism for helimagnetism is likely distinct from that in Sr(Co$_{1-x}$Ni$_x$)$_2$As$_2$. Here, a classical Heisenberg model with a frustrated nearest-neighboring (NN) $J_1$ and next-nearest-neighboring (NNN) $J_2$ between basal planes can produce helical order \cite{Johnston2017,Nagamiya1968}. Such a  mechanism of helical order for centro-symmetric materials has been proposed many years ago by incorporating exchange frustration\cite{Heli1}. However, there are difficulties in applying this model to Sr(Co$_{1-x}$Ni$_x$)$_2$As$_2$. First, the effective spin $S$ derived from $\mu_{eff}$ = $g \sqrt{S(S+1)}$ is only about 1/2, making the validity of classical Heisenberg model ($S \rightarrow \infty$) questionable. Second, according to the theory, the helical arrangement has a $q$ given by $\cos(cq) = -J_1/4J_2$ in which $c$ is the inter-layer distance. Therefore, the nearly $90^{\circ}$ angle between magnetic moments in adjacent layers with $q \sim \frac{\pi}{2c}$ leads to $\cos{cq} \sim \cos(\pi/2) = 0$ and thus an almost vanishing NN exchange coupling $J_1$ which appears to be unrealistic. Higher order exchange couplings such as $J_3$ may help stabilize the helical state but the phase space \cite{Nagamiya1962} for a stable helix with a rotation of $\sim\pi/2$ angle between adjacent spins is still limited. Third, the RKKY interaction may be responsible for the evolution of magnetism from CaCo$_2$As$_2$ to BaCo$_2$As$_2$, since the different sizes of intercalated ions can tune the inter-layer Co-Co distance and thus the magnitude of exchange interactions. However, the origin of the helimagnetism along the $c$ axis is still ambiguous because an unrealistically large $k_F$ ($\sim \frac{\pi}{c}$) is required to make $J_1$ nearly zero via the spatial oscillation of the RKKY interaction.

On the other hand, quantum fluctuations in a paramagnet near a FM instability can make significant contributions to the free energy and stabilize unusual magnetic order \cite{Uhlarz2004,Conduit,Green}. In Sr(Co$_{1-x}$Ni$_x$)$_2$As$_2$, the $d_{x^2-y^2}$ band is flat along both the in-plane $\Gamma-M$ and out-of-plane $\Gamma-Z$ directions. The massive quantum particle-hole excitations can strongly affect the flat band dispersion at the mean-field level and, as a consequence, a helical magnetic order with certain momenta can be induced. However, the presence of flat band requires physics beyond the conventional itinerant spin density wave (SDW) theory with Fermi surfaces nesting, but may be a realization of the \textit{quantum-order-by-disorder} mechanism.

In conclusion, we discovered a novel helical magnetic order in Sr(Co$_{1-x}$Ni$_2$)$_2$As$_2$ and established the phase diagram with systematic studies of the magnetic susceptibility. By combining neutron diffraction and ARPES experiments, we have established a close connection between the $d_{x^2-y^2}$ flat band and the helical magnetic order. Based on these results, we suggest that the helical magnetic order in Sr(Co$_{1-x}$Ni$_x$)$_2$As$_2$ is driven by quantum fluctuations and the system is likely in the vicinity of at least one quantum critical point.

Note added:  Recently, a paper studying\cite{SCNA} on the same material as the present work was posted on arXiv. The main experimental results of the two studies are consistent with each other.

\section{Acknowledgments}

We are grateful to Dr. D.L. Gong and Dr. L.Y. Xing for the assistance with data analysis and thank Dr. R.Y. Jin, Dr. J.D. Zhang, and Dr. Ilya Vekhter for helpful discussions. Y.L. would like to thank Dr. Shiliang Li and Dr. Huiqian Luo for providing lab equipment for sample growth and Beijing Normal University for sample characterization during his stay in China. Single crystal growth and the work for neutron scattering were supported by the U.S. NSF DMR-1700081 and the Robert A. Welch Foundation Grant No. C-1839 (P.D.). The MPMS measurements, data analysis, and manuscript preparation were supported by the U.S. Department of Energy under EPSCoR Grant No. DE-SC0012432 with additional support from the Louisiana Board of Regents. The ARPES experiment was performed on the Dreamline beamline, Shanghai Synchrotron Radiation Facility and supported by Ministry of Science and Technology of China (2016YFA0401002) and CAS Pioneer Hundred Talents Program (type C). This research used resources at the High Flux Isotope Reactor, a DOE Office of Science User Facility operated by the Oak Ridge National Laboratory.

{}

\end{document}